\documentclass[aps,preprint,tightenlines,superscriptaddress,showpacs]{revtex4}
\usepackage{epsfig}
\begin{document}
\title{Towards an understanding of heavy baryon spectroscopy}

\author{A. Valcarce}
\affiliation{Departamento de F\'\i sica Fundamental, 
Universidad de Salamanca, E-37008 Salamanca, Spain}
\author{H. Garcilazo}
\affiliation{Escuela Superior de F\'\i sica y Matem\'aticas,
Instituto Polit\'ecnico Nacional, Edificio 9, 07738 M\'exico D.F., Mexico}
\author{J. Vijande}
\affiliation{Departamento de F\' \i sica Te\'orica e IFIC,
Universidad de Valencia - CSIC, E-46100 Burjassot, Valencia, Spain}

\date{\today}

\begin{abstract}
The recent observation at CDF and D0 of $\Sigma_b$, $\Sigma^*_b$ and 
$\Xi_b$ baryons opens the door to the advent of new states in the
bottom baryon sector. The states measured provide sufficient constraints 
to fix the parameters of phenomenological models. One may therefore 
consistently predict the full bottom baryon spectra. 
For this purpose we have solved exactly the three-quark problem by means
of the Faddeev method in momentum space. We consider our
guidance may help experimentalists in the search
for new bottom baryons and their findings will
help in constraining further the phenomenological models.
We identify particular states whose masses may allow to
discriminate between the dynamics for the light-quark pairs
predicted by different phenomenological models. 
Within the same framework we also present results for
charmed, doubly charmed, and doubly bottom baryons.
Our results provide a restricted possible assignment 
of quantum numbers to recently reported charmed baryon states.
Some of them are perfectly described by 
$D-$wave excitations with $J^P=5/2^+$, as the
$\Lambda_c (2880)$, $\Xi_c(3055)$,
and $\Xi_c(3123)$. 
\end{abstract}

\pacs{12.39.Jh, 12.39.Pn, 14.20.-c}
\maketitle

\section{Introduction}

The last year has witnessed an amazing experimental progress in the
identification of new bottom baryon states. Last June CDF~\cite{Aal07a}
reported the first observation of the 
$\Sigma^\pm_b$, $uub$ and $ddb$ states with $J^P=1/2^+$,
and $\Sigma^{\star \pm}_b$, $uub$ and $ddb$ states with $J^P=3/2^+$.
The observed values were
\begin{eqnarray}
M(\Sigma^-_b) & = &  5815.2 \pm 1.0 ({\rm stat}) \pm 1.7 ({\rm syst}) \, {\rm MeV/c^2} \, , \cr
M(\Sigma^+_b) & = &  5807.8^{+2.0}_{-2.2} ({\rm stat}) \pm 1.7 ({\rm syst}) \, {\rm MeV/c^2}  \, ;
\end{eqnarray}
and
\begin{eqnarray}
M(\Sigma^{\star -}_b) & = &  5836.4^{+2.0}_{-1.8} ({\rm stat})\,\, ^{+1.8}_{-1.7} ({\rm syst}) \, {\rm MeV/c^2} \, , \cr
M(\Sigma^{\star +}_b) & = &  5829.0^{+1.6}_{-1.8} ({\rm stat})\,\, ^{+1.7}_{-1.8} ({\rm syst}) \, {\rm MeV/c^2}  \, .
\end{eqnarray}
During June and July CDF~\cite{Aal07b} and D0~\cite{Aba07} reported the observation
of the baryon $\Xi^-_b$, $dsb$ state with $J^P=1/2^+$,
\begin{eqnarray}
M(\Xi^{-}_b) & = &  5792.9 \pm 2.5 ({\rm stat})\,\, \pm 1.7 ({\rm syst}) \, {\rm MeV/c^2} \,\, ({\rm CDF})\, , \cr
M(\Xi^{-}_b) & = &  5774 \pm 11 ({\rm stat})\,\, \pm 15 ({\rm syst}) \, {\rm MeV/c^2} \,\, ({\rm D0})\, . 
\end{eqnarray}
The number of known bottom baryons increased from one to four
over a few months, determining for the first time the hyperfine splitting
in the bottom sector.

Heavy hadrons containing a single heavy quark are particularly
interesting. The light degrees of freedom (quarks and gluons)
circle around the nearly static heavy quark. Such a system behaves as
the QCD analogue of the familiar hydrogen bounded by the 
electromagnetic interaction. 
When the heavy quark mass $m_Q \to \infty$, the angular
momentum of the light degrees of freedom is a good 
quantum number. Thus, heavy quark baryons belong to 
either SU(3) antisymmetric $\mathbf{\bar{3}_F}$ or symmetric 
$\mathbf{6_F}$ representations. The spin of the light
diquark is 0 for $\mathbf{\bar{3}_F}$, while it is 1
for $\mathbf{6_F}$. Thus, the spin of the ground
state baryons is $1/2$ for $\mathbf{\bar{3}_F}$, representing
the $\Lambda_b$ and $\Xi_b$ baryons, while it can be
both $1/2$ or $3/2$ for $\mathbf{6_F}$, allocating
$\Sigma_b$, $\Sigma^*_b$, $\Xi'_b$, $\Xi^*_b$, $\Omega_b$
and $\Omega^*_b$, where the star indicates spin $3/2$ states.
Therefore heavy hadrons form doublets.
For example, $\Sigma_b$ and $\Sigma_b^*$ will be degenerate
in the heavy quark limit. Their mass splitting is caused
by the chromomagnetic interaction at the order $1/m_Q$.
These effects can be, for example, taken into account systematically in the
framework of heavy quark effective field theory (HQET).
The mass difference between states belonging to the 
$\mathbf{\bar{3}_F}$ and $\mathbf{6_F}$ representations
do also contain the dynamics of the light diquark 
subsystem, hard to accommodate in any heavy quark mass
expansion. Therefore, exact solutions of the three-body
problem for heavy hadrons are theoretically
desirable because they will serve to test the reliability   
of approximate techniques, that would only
be exact in the infinite heavy-quark mass limit, as could
be heavy quark mass expansions, variational calculations, or quark-diquark
approximations. 

Heavy baryons, charmed and/or bottom,
have been the matter of study during the last two
decades~\cite{Cop79,Mal80,Ric83,Cap86,Ron95,Sil96,Jen96,Bow96,
Mat02,Ebe05,Bag92,Wan02,Kar07,Liu07,Rob07,Ebe08}.
After the discovery of the first charmed baryons, several theoretical 
works~\cite{Cop79,Mal80,Ric83} based on potential models developed 
for the light baryon or meson spectra started analyzing properties of
the observed and expected states. Later on, 
Capstick and Isgur~\cite{Cap86} studied heavy baryon systems in a relativized
quark potential model applying a variational approach
to obtain the mass eigenvalues and bound state wave functions
by using a harmonic oscillator basis. Roncaglia {\it et al.}~\cite{Ron95}
predicted the masses of baryons containing one or two heavy quarks using the
Feynman-Hellmann theorem and semiempirical mass formulas.
Silvestre-Brac~\cite{Sil96} studied ground state charmed and bottom baryons using 
Faddeev equations in configuration space. Excited states
were studied by diagonalization in a harmonic oscillator basis up to
eight quanta.
Jenkins~\cite{Jen96} studied heavy baryon masses in a combined expansion
in $1/m_Q$, $1/N_c$, and SU(3) flavor symmetry breaking.
Bowler {\it et al.}~\cite{Bow96} made an exploratory study 
using lattice techniques to predict charmed and bottom baryons.
Mathur {\it et al.}~\cite{Mat02} gave a more precise prediction of
the masses of charmed and bottom baryons from quenched lattice QCD.
Ebert {\it et al.}~\cite{Ebe05} calculated the masses of ground state heavy baryons with the
relativistic quark-diquark approximation. QCD sum rule has been also
applied to study heavy baryon masses~\cite{Bag92,Wan02}.                  
Stimulated by the recent experimental progress, there have been several theoretical papers
on the masses of $\Sigma_b$, $\Sigma_b^*$ and $\Xi_b$ or the full
bottom baryon spectra using a 
perturbative treatment of the hyperfine interaction in the quark
model~\cite{Kar07}, heavy quark effective field 
theory~\cite{Liu07}, a variational 
calculation in a harmonic oscillator expansion~\cite{Rob07},
and a relativistic quark-diquark approximation~\cite{Ebe08}.

While the mass of heavy baryons is measured 
as part of the discovery process,
no spin or parity quantum numbers of a given state have been 
measured experimentally, but they are assigned based on
quark model expectations. Such properties can only be extracted by
studying angular distributions of the particle decays, that
are available only for the lightest and most abundant species.
For excited heavy baryons the data set
are typically one order of magnitude smaller than for 
heavy mesons and therefore the
knowledge of radially and orbitally 
excited states is very much limited.
Unlike the heavy mesons there are no resonant production
mechanisms and thus heavy baryons can only be obtained by
continuum production, where cross sections are small.
As a consequence the $B$ factories have been the main
source of these baryons.  
Therefore, a powerful guideline for assigning quantum numbers 
to new states or to indicate new states to look for is 
required by experiment. We do understand ground state heavy
quark baryons, both in the quark model and in the lattice QCD.
The main issue is therefore to determine quantum numbers of excited
states. Here, a coherent theoretical and experimental effort is
required.

Apart from CDF and D0 data, putting into operation the Large Hadron
Collider (LHC) will provide us with data on masses of excited
bottom baryons. Therefore the calculation of the mass spectra 
of excited heavy baryons turns out to be a really actual problem.
Here we consider the exact calculation of ground states, spin, radial and
orbital excitations of bottom baryons in a model constrained to
reproduce the new recent experimental data. These new experimental
data give rise to a spin splitting of the order of 25 MeV, 
much smaller that previous experimental 
expectations, of the order of 50 MeV~\cite{Bow96,Fei95}.
Using the same phenomenological model we also calculate the
charmed baryon spectra showing the nice agreement of our
predictions with recently measured states,
what will allow to assign their
spin and parity quantum numbers. We will finally 
consistently present our predictions
for doubly bottom and charmed baryons. 

\section{Formalism and results}

Nowadays, we have to our disposal {\it realistic} quark models accounting
for most part of the one- and two-body low-energy hadron phenomenology.
The ambitious project of a simultaneous description 
of the baryon-baryon interaction and the baryon (and meson) 
spectra has been undertaken by the constituent quark 
model of Ref. \cite{Val05}. The success in describing the 
properties of the strange and non-strange one and two-hadron
systems encouraged its use as a guideline in order to assign parity 
and spin quantum numbers to already determined heavy baryon
states as well as to predict still non-observed resonances~\cite{Gar07}. 
The constituent quark model used considers
perturbative (one-gluon exchange) and nonperturbative (confinement and
chiral symmetry breaking) aspects of QCD, ending up with
a quark-quark interaction of the form 
\begin{equation} 
V_{q_iq_j}=\left\{ \begin{array}{ll} 
q_iq_j=nn/sn\Rightarrow V_{CON}+V_{OGE}+V_{\chi} &  \\ 
q_iq_j=cn/cs/bn/bs/cc/bb \Rightarrow V_{CON}+V_{OGE} & 
\end{array} \right.\,,
\label{pot}
\end{equation}
where $V_{CON}$ stands for a confining interaction, $V_{OGE}$ for
a one-gluon exchange potential, and $V_{\chi}$ for scalar and
pseudoscalar boson exchanges. For heavy quarks ($c$ or $b$) chiral symmetry is
explicitly broken and boson exchanges do not contribute.
For the explicit expressions of the interacting potential and a more
detailed discussion of the model we refer the reader to Refs. \cite{Vij04,Val05b}.
For the sake of completeness we resume the parameters of the model in Table \ref{t1}.
The results have been obtained by solving exactly the 
Schr\"odinger equation by the Faddeev method in momentum space~\cite{Val05}.

The recent observation of ground state $\Sigma_b$ and $\Sigma^*_b$ baryons 
provides with all necessary ingredients to fix the model parameters and therefore
make univocal predictions for all remaining bottom baryon states: spin, radial
and orbital excitations. 
Once the model is fixed we can also derive the spectra
for doubly charmed and doubly bottom baryons. Besides we can revisit
the charmed baryon sector, centering our attention in some 
states recently reported in an attempt to help in the assignment of
spin and parity quantum numbers. Finally, our results 
will allow us to check equal spacing rules derived from heavy quark symmetry
and chiral symmetry. 

Our results for bottom baryons are shown in Table \ref{t2} compared to 
experiment and other theoretical approaches. 
We present our predictions for spin, radial and orbital excitations. 
All known experimental data are nicely described. 
Such a remarkable agreement and the
exact method used to solve the three-body problem make our
predictions highly valuable as a guideline to experimentalists.
They should also serve to guide theoretical calculations using
approximate methods.

As compared to other results
in the literature we see an overall agreement for the low-lying
states both with the quark-diquark approximation of 
Ref.~\cite{Ebe08} and the variational calculation in a
harmonic oscillator basis of Ref.~\cite{Rob07}. 
Some differences appear for the excited states that we will
analyze in the following and that could be
either due to the interacting potential or to the method used
to solve the three-body problem. 
The relativistic quark-diquark approximation of Ref.~\cite{Ebe08} 
predicts a larger radial excitation
for negative parity states (except for the $\Xi_b$ baryon)
as compared to any other result in the literature. 
We do not see any explanation for this result.
The relativistic quark-diquark approximation
and the harmonic oscillator variational method predict a lower
$3/2^+$ excited state for the $\Lambda_b$ baryon.
Such result can be easily understood by looking at Table~\ref{t7},
where it is made manifest the influence  of the pseudoscalar 
interaction between the light quarks on the 
$\Lambda_b(1/2^+)$ ground state, diminishing its mass by 
200 MeV. If this attraction would not be present for the $\Lambda_b(1/2^+)$, the 
$\Lambda_b(3/2^+)$ would be lower in mass as reported
in Refs.~\cite{Rob07,Ebe08} (a similar effect will be 
observed in the charmed baryon spectra).
Thus, the measurement and identification of the 
$\Lambda_b(3/2^+)$ is a relevant feature that will help to
clarify the nature of the interaction between the light
quarks in heavy baryon spectroscopy, determining the need
of pseudoscalar forces consequence of the spontaneous
chiral symmetry breaking in the light flavor sector.

Let us revise our results in connection with the interacting
potential used. 
The first ingredient of any quark potential model is confinement. 
Confinement is supposed to be flavor independent and therefore it should be 
fixed in one flavor sector once for all. As mentioned above, 
knowledge of orbital and radial excited states is very much
limited for heavy baryons. Thus, guidance for the confinement strength should 
be taken from the light baryon sector.
The non-strange baryon sector is the best known one from the spectroscopic 
point of view. However it is delicate to fix the confinement strength due to
the particular nature of the radial excitation, the Roper resonance.
In Refs.~\cite{Gar03} it has been shown the sensitivity of the 
Roper resonance to relativistic kinematics, justifying 
the use of negative parity states to fix the confinement 
strength when working in a non-relativistic framework. 
The strength of confinement quoted in Table~\ref{t1} gives a good 
overall agreement with the $N$ and $\Delta$ spectra.

Once the confinement strength has been fixed in the light flavor
sector, we note that the radial excitation of $1/2^+$ bottom baryons
is predicted around 450 MeV above the ground state.
The only exception is the
$\Xi_b(1/2^+)$ with an excitation energy, $1/2^+ - 1/2'^+$, around
140 MeV. This resonance is not indeed a radial excitation.
The ground state corresponds to a $us$ pair in a dominant singlet 
spin state while the excited state corresponds to the same
pair in a dominant triplet spin state. These two levels are often
denoted in the literature as $\Xi_b(1/2^+)$ and
$\Xi'_b(1/2^+)$, the same notation we will use in this work. 
As can be seen in Table~\ref{t2} their
radial excitations also appear 
450 MeV above the corresponding ground state.
At difference of other calculations in the literature
where the $\Xi_b$ and $\Xi'_b$ baryons are pure scalar or
vector light diquark states~\cite{Rob07,Ebe08} (mixed
is some cases perturbatively), our calculation includes
all possible channels contributing to each state.
The similar results obtained indicate a small 
admixture of scalar and vector diquarks in nature.
We consider very important that the analysis of the
different flavor sectors is done in terms of the same
flavor-independent forces to obtain conclusions about
the rest of the dynamical model. This is not often
mentioned in the literature.

Our screened confining potential would
give rise at short range to a linear potential with a strength 
of about 600 MeV fm$^{-1}$. This value is very close to the 
string tension obtained in Ref.~\cite{Bal05} from the bottomonium
spectrum, $\sqrt{\kappa}=$ 440$-$480 MeV, that
would translate into a linear confinement strength of the same
order, around 500$-$600 MeV fm$^{-1}$. In contrast,
for example, Ref.~\cite{Rob07} uses a smaller 
confining strength of the order of 400 MeV fm$^{-1}$. 
Maybe this is one of the reasons why they use a negligible
coulomb strength.

Being the confinement strength determined in the light baryon sector,
heavy baryons are ideal systems
to study the flavor dependence of the spin splitting, mass
difference between $\Sigma_b(3/2^+)\equiv \Sigma^*_b$ 
and $\Sigma_b(1/2^+)\equiv \Sigma_b$.
Such systems present, on one hand, the dynamics of the two-light
quarks involving potentials coming from the chiral part
of the interaction and, on the other hand, the dynamics of heavy-light
subsystems. The dynamics of the light diquark subsystem is
fully determined in studying the light-baryon spectra,
and therefore its contribution to the spin splitting.
Thus, the remaining spin splitting must be due to
the interaction between the heavy and the light quarks.
The recent measurement of $\Sigma_ b$ and $\Sigma^*_b$ states
determines in a unique way the color-magnetic interaction
between the light and the $b$ quark, allowing a parameter
free prediction of all other states of the bottom spectra. 
A relevant conclusion of the study of the spin splitting for
bottom baryons is that while
for similar mass quark pairs one can use 
for the regularization parameter of the color-magnetic
one-gluon exchange interaction, $r_0$,
a formula depending on the reduced mass of the system
(this has been proved in the past~\cite{Sil96,Val05b}), however, 
when the masses of the interacting quarks 
are quite different (the case of heavy baryons), 
a reduced mass based formula is not adequate.
Such a formula will give the same result, for example, for a
light-charm than for a light-bottom pair. As the color-magnetic
term of the one-gluon exchange interaction depends on the inverse of
the product of the masses of the interacting quarks, such a
potential will be strongly reduced for the heavier pair
producing a too small spin-splitting.

We are thus led to the interplay between the pseudoscalar
and the one-gluon exchange interactions, a key problem 
for both the baryon spectra and the two-nucleon system~\cite{Nak00}. 
This can be illustrated noting that while 
the $\Sigma_i(3/2^+)-\Lambda_i(1/2^+)$ 
mass difference varies slowly from the strange to the bottom sector,
the $\Sigma_i(3/2^+)-\Sigma_i(1/2^+)$ mass difference varies very fast
(see Table~\ref{t5}). As discussed in the introduction,
the former is a mass difference between members of the 
$\mathbf{\bar{3}_F}$ and $\mathbf{6_F}$ SU(3) representations
and therefore it presents contributions
from the pseudoscalar and one-gluon exchange forces (see
columns $V_1$ and $V_3$ of Table III in Ref.~\cite{Val05b}).
However, the latter is a mass difference between members of
the same representation, $\mathbf{6_F}$, and it is therefore
uniquely due to the one-gluon exchange interaction between
the light diquark and the heavy quark. 
This can be easily understood by explicit construction
of the spin-flavor wave-function. In
the case of $\Lambda$ baryons the two light quarks are in a 
flavor antisymmetric spin $0$
state, the pseudoscalar and the one-gluon exchange forces being
both attractive. For $\Sigma$ baryons they are in a flavor
symmetric spin $1$ state. The pseudoscalar force, being still
attractive, is suppressed by one order of magnitude due to the
expectation value of the $(\vec \sigma_i \cdot \vec \sigma_j)
(\vec \lambda_i \cdot \vec \lambda_j)$ operator~\cite{Val05b}
and the one-gluon exchange between the two light quarks
becomes repulsive. Therefore, the attraction is provided by
the interaction between the light diquark and the heavy
quark, which for heavy quarks $c$ or $b$ is given only
by the one-gluon exchange potential. 

The above discussion is explicitly illustrated in Table \ref{t7},
where we have calculated the mass of $\Sigma_i$ and $\Lambda_i$ ($i=s$ or $b$) 
baryons with and without the
pseudoscalar exchange contribution. As can be seen the effect of the 
pseudoscalar interaction between the two-light quarks is 
approximately the same independently of the third quark.
As the mass difference between the $\Sigma_i(3/2^+)$ and
$\Sigma_i(1/2^+)$ states decreases when increasing the mass of the baryon,
being almost constant the effect of the one pion-exchange, the 
remaining mass difference has to be
accounted for by the one-gluon exchange (mass difference between states
belonging to the $\mathbf{6_F}$ representation). 
This rules out any ad hoc recipe for the relative strength of both
potentials, what would be in any manner consistent with experiment,
and it also reinforces the importance
of constraining models for the baryon spectra in the 
widest possible set of experimental data. Thus, Table~\ref{t7}
shows how for heavy quark baryons
the dynamics of any two-particle subsystem is not affected
by the nature of the third particle. As a consequence, the regularization 
parameter of the one-gluon exchange potential 
should depend on the interacting pair, independently of the
baryon the pair belongs to. The values of $r_0$ reproducing
the experimental data are quoted in Table~\ref{t6}. They obey a formula
depending on the product of the masses of the interacting quarks that can
be represented by
$r_0^{q_iq_j}= A \mu \left( m_{q_i} m_{q_j} \right)^{-3/2}$, where
$A$ is a constant and $\mu$ the reduced mass of the interacting quarks.
While working with almost equal or not much different masses this law can be
easily replaced by a formula depending on some inverse power of the mass (or
reduced mass) of the pair as obtained in Ref.~\cite{Vij04}, but this is not any more the
case for quarks with very different masses, like those present
in heavy baryons. This is one of the reasons why these systems constitute
an excellent laboratory for testing low-energy QCD realizations.

Based on our exact method for the solution of the three-body 
problem, let us analyze the predictions derived from 
heavy-quark symmetry (HQS) and chiral symmetry 
combined together in order to describe the soft hadronic
interactions of hadrons containing a heavy quark~\cite{Wis92}. In the
limit of the heavy quark mass $m_Q \to \infty$, HQS predicts that all
states in the $\mathbf{6_F}$ SU(3) representation (those where the light 
degrees of freedom are in a spin 1 state) would be degenerate.
If one considers HQS and lowest order SU(3) breaking~\cite{Sav95} the masses
of heavy baryons obey an equal spacing rule,
similar to the one that arises in the decuplet of light $J^P=3/2^+$
baryons, it reads
\begin{eqnarray}
J=1/2 \,\, : \,\, M_{\Sigma_b} \, + \, M_{\Omega_b} \, & = & \, 2\, M_{\Xi'_b} \nonumber \\
J=3/2 \,\, : \,\, M_{\Sigma^*_b} \, + \, M_{\Omega^*_b} \, & = & \, 2\, M_{\Xi^*_b} \,.
\label{spa}
\end{eqnarray}
This equal spacing rule also holds for the hyperfine splittings
\begin{equation}
\delta_{\Sigma_b} \, + \, \delta_{\Omega_b} \, = \, 2\, \delta_{\Xi_b} \, ,
\label{spa2}
\end{equation}
where $\delta_{\Sigma_b}\, =\, M_{\Sigma^*_b}\, - \, M_{\Sigma_b}$,
$\delta_{\Xi_b}\, =\, M_{\Xi^*_b}\, - \, M_{\Xi'_b}$, and
$\delta_{\Omega_b}\, =\, M_{\Omega^*_b}\, - \, M_{\Omega_b}$.
The latter relation is expected to be more accurate than 
Eq.~(\ref{spa})~\cite{Jen96} and it is exactly fulfilled
by our results as it is shown in Table~\ref{t11}.
Combining Eqs.~(\ref{spa}) and (\ref{spa2}), one arrives
to the approximate relations,
\begin{eqnarray}
\Xi'_b(1/2^+) - \Sigma_b(1/2^+) &=&
\Omega_b(1/2^+) - \Xi'_b(1/2^+)= \nonumber \\
= \Xi_b(3/2^+) - \Sigma_b(3/2^+) &=&
\Omega_b(3/2^+) - \Xi_b(3/2^+) \, .
\label{eqs}
\end{eqnarray}
These relations are satisfied by experimental data 
in the case of charmed baryons with differences of the order of 
10 MeV, what gives an idea of the breaking of the SU(3) flavor
symmetry. These predictions are clearly sustained by our model 
to the same precision as can be checked in Table~\ref{t10}. Therefore 
our dynamical model incorporates the features of broken SU(3) flavor 
symmetry and heavy-quark expansion of QCD in a reasonable way.
Let us note that one-gluon exchange based models do almost 
satisfy exactly Eq.~(\ref{eqs})
(results of Refs.~\cite{Sil96} and \cite{Rob07}
in Table~\ref{t10}) probably due to the absence of SU(3) flavor symmetry
breaking interactions. It is also interesting to note that lattice calculations
based on HQS fulfill exactly Eq.~(\ref{eqs}) for charmed baryons~\cite{Bow96} while
the disagreement is very large for bottom baryons (results of Ref.~\cite{Bow96}
in Table~\ref{t10}). 

To have the widest set of predictions within the same model we
have calculated the charmed baryon spectra.
The results are shown in Table~\ref{t12}. In this case
the masses of several ground and excited states are known. They fit nicely
within our results based on the quantum numbers assigned by the PDG using
quark model arguments. There are some excited charmed baryons that it is
not even known if they are excitations of the $\Lambda_c$ or $\Sigma_c$.
The first one is a resonance at 2765 MeV reported by CLEO in Ref.~\cite{Art01}, 
that could be either a $\Lambda_c$ or a $\Sigma_c$ baryon. The original
reference suggested the possibility of a $\Sigma_c$ state with $J^P=1/2^-$.
For this state our calculation shows a perfect agreement with the suggested
nature and quantum numbers, although one could not discard this state being
the first radial excitation $2S$ of the $\Lambda_c$ with $J^P=1/2^+$, as 
has been suggested in Refs.~\cite{Ebe05,Rob07}. 
The $\Lambda_c(2880)$ is perfectly described by the two possibilities
suggested by experiment: either an orbital excitation $2P$ 
of the $\Lambda_c$ with $J^P=1/2^-$, as conjectured by CLEO
in Ref.~\cite{Art01} due to the observation of decay via $\Sigma_c \pi$
but not via $\Sigma_c^* \pi$, or an orbital excitation 
$1D$ of the $\Lambda_c$ with $J^P=5/2^+$, in agreement with the 
recent spin assignment by Belle based on the analysis of angular distributions
in the decays $\Lambda_c(2880)^+ \to \Sigma_c(2455)^{0,++} \pi^{+,-}$%
~\cite{Abe07}. Our model also predicts the resonance at 2940 MeV recently
reported by BaBar in Ref.~\cite{Aub07} being the
first radial excitation $2S$ of the $\Sigma_c$ with $J^P=3/2^+$. Finally,
the $\Sigma_c(2800)$ may correspond to the second state of the lowest
$P$ wave multiplet with $J^P=3/2^-$, very close to its $1/2^-$
partner at 2765 MeV. 
A number of new $\Xi_c$ and $\Xi'_c$ states have been also discovered recently.
Two resonances at 3055 and 3123 MeV have been reported
by BaBar in Ref.~\cite{Aubb07}. They fit into the doublet of orbital
excited states $2D$ with $J^P=5/2^+$, the first one with the light
diquark in a spin 0 state, $\Xi_c$, and the second in a spin 1 state, $\Xi'_c$.
For the second resonance one cannot discard
the first radial excitation $2S$ of the $\Xi_c$ (light diquark
in a spin 0 state) with $J^P=1/2^+$ as suggested in Ref.~\cite{Rob07}. 
Belle, in Ref.~\cite{Miz07}, has reported two resonances
at 2980 and 3076, while the first one may correspond 
to the first radial excitation $2P$ of the $\Xi_c$ with $J^P=1/2^-$,
the second clearly corresponds to the first radial excitation
$2S$ of the $\Xi_c$ with $J^P=3/2^+$.
As can be seen all known experimental states
fit nicely into the description of our model not leaving too many
possibilities open for the assigned quantum numbers as we have resumed
in Table~\ref{t30}.

Finally, we can easily extend our predictions to doubly bottom and charmed
baryons. A few years ago the SELEX Collaboration~\cite{Och05}
reported the discovery of a baryon with a mass of 3519 GeV that they
concluded could be a doubly charmed $\Xi_{cc}$ state. The attempts to
confirm such discovery by BaBar~\cite{Aub06}, Belle~\cite{Chi06},
and FOCUS~\cite{Rat03} Collaborations have failed. Potential models
based on chromomagnetic interactions predict for this state larger
masses~\cite{Rob07}. In our case we can make parameter free predictions
for ground states as well as for spin, orbital and radial excitations.
The ground state is found to be at 3579 MeV, far below the result of
Ref.~\cite{Rob07} and in perfect agreement with lattice nonrelativistic
QCD~\cite{Mat02},
but still a little bit higher than the non-confirmed SELEX result. It is therefore
a challenge for experimentalists to confirm or to find the ground state
of doubly charmed and bottom baryons. 

The combined study of $Qqq$ and $QQq$ systems, where $Q$ stands for a heavy
$c$ or $b$ quark and $q$ for a light $u$, $d$, or $s$ quark, will also provide 
some hints to learn about the basic dynamics governing the interaction
between light quarks. The interaction between
pairs of quarks containing a heavy quark $Q$ is driven by the 
perturbative one-gluon exchange. This has been demonstrated by
quenched and unquenched lattice QCD calculations~\cite{Bal05}.
The important issue of the simultaneous study of these two types
of heavy baryons, is the presence
in one of them of a pair of light quarks.
As explained above, for the $Qqq$
system the mass difference between members of the  
$\mathbf{6_F}$ SU(3) representation comes determined only
by the perturbative one-gluon exchange, whether between members
of the $\mathbf{6_F}$ and $\mathbf{\bar{3}_F}$ representations
it presents contributions from the one-gluon exchange and also
possible pseudoscalar exchanges. If the latter mass difference
would be attributed only to the one-gluon exchange (this would be the case
of models based only on the perturbative one-gluon exchange), it will be strengthened
as compared to models considering pseudoscalar potentials at the 
level of quarks, where a weaker one-gluon exchange will play the role.
When moving to the $QQq$ systems only one-gluon exchange interactions
between the quarks will survive, with the strength determined in the
$Qqq$ sector, where we have experimental data. This will give rise
to larger masses for the ground states, due to the more attractive 
one-gluon exchange potential in the $Qqq$ sector, what requires
larger constituent quark masses to reproduce the experimental data.
This could be the reason for the larger masses of ground state doubly
heavy baryons obtained with gluon-based interacting 
potentials~\cite{Cap86,Rob07}. 

Therefore, among the baryons with two heavy quarks the first question
to be settled is where do exactly these states lie. In case such low 
masses as those reported by SELEX were
confirmed theorists will have a challenge to accommodate this state
into the nice description of charmed and bottom baryons. 
Their exact mass may help in discriminating between the dynamics of
the light degrees of freedom of the different models.
In any case, the excited spectra of doubly charmed and bottom 
baryons do not depend much on the mass of the heavy quarks, 
and therefore the predicted excited
spectra should serve as a guideline for potential future experiments
looking for such states.   
Our results for the ground and excited spectra are resumed in Tables~\ref{t19} 
and~\ref{t20} compared to those of Refs.~\cite{Ron95,Sil96,Mat02,Rob07}.
As can be seen the radial excitations of Ref.~\cite{Rob07}
are lower than in our model, due to the small confining 
strength used. We also note some unexpected results in
Ref.~\cite{Rob07} as the reverse of the hierarchy in 
the spin splitting between $\Xi_{bb}$ and $\Omega_{bb}$
compared to $\Xi_{cc}$ and $\Omega_{cc}$, what could be
a misprint in this reference.

\section{Summary}
\label{sec5}

We have used a constituent quark model incorporating the basic properties
of QCD to study the bottom baryon spectra. 
Consistency with the light baryon spectra and the new experimental data
reported by CDF allow to fix all model parameters. 
The model takes into account the most
important QCD nonperturbative effects: chiral symmetry breaking
and confinement as dictated by unquenched lattice QCD. It also
considers QCD perturbative effects trough a flavor dependent one-gluon exchange
potential. We make a parameter free prediction
of the spin, orbital and radial excitations. 
We have predicted the spectra of doubly bottom and charmed baryons.
We have also revisited the charmed baryon spectra finding a nice 
agreement with the recently reported data what allowed to make 
a restricted assignment of their spin and parity quantum numbers.

Our results have been obtained by solving exactly 
the three-body problem by means of the
Faddeev method in momentum space. In spite of the huge
computer time needed to obtain the set of results presented
in this work, such effort should be highly
valuable both from the theoretical and experimental points of 
view. Theoretically, it should be a powerful tool
for testing different approximate methods to solve the 
three-body problem in the large mass limit for one or two
of the components. Experimentally, the remarkable 
agreement with known experimental data make our
predictions highly valuable as a guideline to experimentalists.

The flavor independence of the confining
potential has been used to describe all flavor sectors.
We have identified particular states of single heavy baryons
whose masses will be clearly
different depending on the particular dynamics governing
the interaction between the two light quarks. The measurement
and identification of the $\Lambda_i(3/2^+)$ ($i=c$ or $b$) 
will provide enough information to distinguish between the
two alternatives for the light quark dynamics: only gluon 
exchanges or gluon supplemented by pseudoscalar forces.
In our description we notice a key interplay
between pseudoscalar and one-gluon exchange forces, already observed
for the light baryons, that may constitute a basic ingredient for the 
description of heavy baryons.
The final spectra results from a subtle but physically meaningful
balance between different spin-dependent forces. The baryon spectra make 
manifest the presence of two different sources of spin-dependent
forces that can be very well mimic by the operatorial dependence
generated by the pseudoscalar and one-gluon exchange potentials.

Heavy baryons constitute an extremely interesting problem
joining the dynamics of light-light and heavy-light subsystems
in an amazing manner. While the mass difference between members
of the same SU(3) configuration, either 
$\mathbf{\bar{3}_F}$ or $\mathbf{6_F}$, is determined
by the perturbative one-gluon exchange, the mass difference
between members of different representations comes mainly
determined by the dynamics of the light diquark, and should
therefore be determined in consistency with the light 
baryon spectra. There is therefore a remnant effect
of pseudoscalar forces in the two-light quark subsystem. 
Models based
only on boson exchanges cannot explain the
dynamics of heavy baryons, but it becomes also difficult for models
based only on gluon exchanges, if consistency between light and
heavy baryons is asked for. One-gluon exchange models would reduce
the problem to a two-body problem controlled by the dynamics of the
heaviest subsystem, and we find evidences in the spectra for 
contributions of both subsystems.

Our results contain the equal mass spacing rules obtained for heavy
baryons by means of heavy quark symmetry and lowest order
SU(3) flavor symmetry breaking to the same accuracy than 
experimental data. The study of the charmed sector shows a
nice agreement with most of the states recently reported
and it allows to predict spin and parity 
quantum numbers for recently measured states.
The experimental confirmation of these assignments
would give further support to the dynamical model used.

We are probably seeing the arrival of a possible understanding
of basic features of quark dynamics in phenomenological models.
The parametrization of the true degrees of freedom of any theory
becomes a challenge that will allow us to advance in the understanding
of low-energy realizations of QCD. Although this has been searched
studying large-orbital angular momenta baryon states or subtle effects
in the light baryon spectra, the combined study of light, heavy and
doubly heavy baryons could be the appropriate laboratory for these
achievements.

\section{acknowledgments}

This work has been partially funded by Ministerio de Ciencia y Tecnolog\'{\i}a
under Contract No. FPA2007-65748, by Junta de Castilla y Le\'{o}n
under Contract No. SA016A07, by the Spanish 
Consolider-Ingenio 2010 Program CPAN (CSD2007-00042), 
and by COFAA-IPN (M\'exico).

\begin{table}[tbp]
\caption{Quark-model parameters.}
\label{t1}
\begin{center}
\begin{tabular}{|cc|ccc|}
\hline
&&$m_u=m_d$ (MeV) & 313 & \\ 
&Quark masses&$m_s$ (MeV)     & 545 & \\ 
&&$m_c$ (MeV) & 1659 & \\ 
&&$m_b$ (MeV) & 5034 & \\ 
\hline
&&$m_{\pi}$ (fm$^{-1}$) & 0.70&\\
&&$m_{\sigma}$ (fm$^{-1}$)& 3.42&\\ 
&&$m_{\eta}$ (fm$^{-1}$)  & 2.77&\\ 
&Boson exchanges&$m_K$ (fm$^{-1}$)       & 2.51&\\ 
&&$\Lambda_{\pi}=\Lambda_{\sigma}$ (fm$^{-1}$) & 4.20 &\\
&&$\Lambda_{\eta}=\Lambda_K$ (fm$^{-1}$) & 5.20&\\ 
&& $g_{ch}^2/(4\pi)$      & 0.54&\\ 
&&$\theta_P(^o)$          & $-$15&\\ 
\hline
&&$a_c$ (MeV)             & 340&\\
&Confinement&$\mu_c$ (fm$^{-1}$)&0.70&\\ 
\hline
&OGE&$r_0$ (fm)     & see Table~\protect\ref{t6}&\\ \hline
\end{tabular}
\end{center}
\end{table}

\begin{table}[tbp]
\caption{Masses of bottom baryons from the present work (CQC)
and other approaches in the literature compared to experimental data (in MeV).
In all cases we quote the central values, in 
Ref.~\protect\cite{Mat02} error bars are of the order of 100 MeV,
40 MeV in Ref.~\protect\cite{Ron95},
125 MeV in Ref.~\protect\cite{Liu07}.}
\label{t2}
\begin{center}
\begin{tabular}{|cc||cccccccccccc|}
\hline
State & $J^P$ & CQC & Exp.  & 
\protect\cite{Cap86} &
\protect\cite{Ron95} &
\protect\cite{Sil96} &
\protect\cite{Jen96} &
\protect\cite{Mat02} &
\protect\cite{Wan02} &
\protect\cite{Kar07} &
\protect\cite{Liu07} &
\protect\cite{Rob07} & 
\protect\cite{Ebe08} \\
\hline
$\Lambda_b$  &  $1/2^+$ & 5624 & 5624 & 5585 & 5620 & 5638 & 5623 & 5672 &      &      & 5637 & 5612 & 5622 \\
             &  $1/2^+$ & 6106 &      & 6045 &      & 6188 &      &      &      &      &      & 6107 & 6086 \\
             &  $1/2^-$ & 5947 &      & 5912 &      & 5978 &      &      &      & 5929 &      & 5939 & 5930 \\
             &  $1/2^-$ & 6245 &      & 6100 &      & 6268 &      &      &      &      &      & 6180 & 6328 \\
             &  $3/2^+$ & 6388 &      & 6145 &      & 6248 &      &      &      &      &      & 6181 & 6189 \\
             &  $3/2^+$ & 6637 &      & 6305 &      & 6488 &      &      &      &      &      & 6401 & 6540 \\ \hline
$\Sigma_b$   &  $1/2^+$ & 5807 & 5808 & 5795 & 5820 & 5845 & 5828 & 5847 & 5790 &      & 5809 & 5833 & 5805 \\
             &  $1/2^+$ & 6247 &      & 6200 &      & 6370 &      &      &      &      &      & 6294 & 6202 \\
             &  $1/2^-$ & 6103 &      & 6070 &      & 6155 &      &      &      &      &      & 6099 & 6108 \\
             &  $1/2^-$ & 6241 &      & 6170 &      & 6245 &      &      &      &      &      &      & 6401 \\
             &  $3/2^+$ & 5829 & 5829 & 5805 & 5850 & 5875 & 5852 & 5871 & 5820 &      & 5835 & 5858 & 5834 \\
             &  $3/2^+$ & 6260 &      & 6250 &      & 6385 &      &      &      &      &      & 6308 & 6222 \\ \hline
$\Xi_b$      &  $1/2^+$ & 5801 & 5793 &      & 5810 & 5806 & 5806 & 5788 &      & 5788 & 5780 & 5844 & 5812 \\
             &  $1/2^+$ & 6258 &      &      &      & 6306 &      &      &      &      &      &      & 6264 \\
             & $1/2'^+$ & 5939 &      &      & 5950 & 5941 & 5950 & 5936 &      &      & 5903 & 5958 & 5937 \\
             & $1/2'^+$ & 6360 &      &      &      & 6416 &      &      &      &      &      &      & 6327 \\
             &  $1/2^-$ & 6109 &      &      &      & 6116 &      &      &      & 6106 &      & 6108 & 6119 \\
             &  $1/2^-$ & 6223 &      &      &      & 6236 &      &      &      &      &      & 6192 & 6238 \\
             &  $3/2^+$ & 5961 &      &      & 5980 & 5971 & 5968 & 5959 &      &      & 5929 & 5982 & 5963 \\
             &  $3/2^+$ & 6373 &      &      &      & 6356 &      &      &      &      &      & 6294 & 6341 \\ \hline
$\Omega_b$   &  $1/2^+$ & 6056 &      &      & 6060 & 6034 & 6061 & 6040 &      & 6052 & 6036 & 6081 & 6065 \\
             &  $1/2^+$ & 6479 &      &      &      & 6504 &      &      &      &      &      & 6472 & 6440 \\
             &  $1/2^-$ & 6340 &      &      &      & 6319 &      &      &      &      &      & 6301 & 6352 \\
             &  $1/2^-$ & 6458 &      &      &      & 6414 &      &      &      &      &      &      & 6624 \\
             &  $3/2^+$ & 6079 &      &      & 6090 & 6069 & 6074 & 6060 & 6060 & 6083 & 6063 & 6102 & 6088 \\
             &  $3/2^+$ & 6493 &      &      &      & 6519 &      &      &      &      &      & 6478 & 6518 \\ \hline
\end{tabular}
\end{center}
\end{table}

\begin{table}[tbp]
\caption{Masses, in MeV, of different bottom baryons with two-light
quarks with (Full) and without ($V_\pi=0$) the contribution of the one-pion exchange
potential. The same results have been extracted from Table III of 
Ref. \protect\cite{Val05b} for strange baryons. $\Delta E$ stands 
for the difference between both results.}
\label{t7}
\begin{center}
\begin{tabular}{|c||ccc|}
\hline
 State & Full & $V_\pi=0$ & $\Delta E$  \\
\hline
$\Sigma_b(1/2^+)$     & 5807 & 5822  & $-$15   \\
$\Sigma_b(3/2^+)$     & 5829 & 5844  & $-$15   \\
$\Lambda_b(1/2^+)$    & 5624 & 5819  & $-$195   \\
$\Lambda_b(3/2^+)$    & 6388 & 6387  & $+$ 1   \\ \hline\hline
State & $V_{CON}+V_{OGE}+V_\pi$ & $V_{CON}+V_{OGE}$ & $\Delta E$  \\
\hline
$\Sigma(1/2^+)$     & 1408 & 1417 & $-$9   \\
$\Sigma(3/2^+)$     & 1454 & 1462 & $-$8   \\
$\Lambda(1/2^+)$    & 1225 & 1405 & $-$180  \\ \hline
\end{tabular}
\end{center}
\end{table}

\begin{table}[tbp]
\caption{Mass difference (in MeV) between $\Sigma_i$ and $\Lambda_i$ states for 
different flavor sectors.}
\label{t5}
\begin{center}
\begin{tabular}{|c||cccccc|}
\hline
Mass difference  & Exp.  & CQC &  
\protect\cite{Sil96} & 
\protect\cite{Ebe05,Ebe08} & 
\protect\cite{Liu07} & 
\protect\cite{Rob07} \\
\hline
$\Sigma(3/2^+)-\Lambda(1/2^+)$     & 269 & 260 & $-$ & $-$ & $-$ & $-$ \\
$\Sigma(3/2^+)-\Sigma(1/2^+)$      & 195 & 169 & $-$ & $-$ &  $-$ & $-$ \\
$\Sigma_c(3/2^+)-\Lambda_c(1/2^+)$ & 232 & 217 & 250 & 221 & 263 & 251 \\
$\Sigma_c(3/2^+)-\Sigma_c(1/2^+)$  &  64 &  67 &  80 &  79 & 123 &  64 \\
$\Sigma_b(3/2^+)-\Lambda_b(1/2^+)$ & 209 & 205 & 237 & 212 & 198 & 246 \\
$\Sigma_b(3/2^+)-\Sigma_b(1/2^+)$  &  22 &  22 &  30 &  29 &  26 &  25 \\ \hline
\end{tabular}
\end{center}
\end{table}

\begin{table}[tbp]
\caption{$r_0^{q_iq_j}$ in fm.}
\label{t6}
\begin{center}
\begin{tabular}{|cccc|}
\hline
& $(q_i,q_j)$ & $r_0^{q_iq_j}$ & \\
\hline
& $(n,n)$     & 0.530 & \\
& $(s,n)$     & 0.269 & \\
& $(n,c)$     & 0.045 & \\
& $(s,c)$     & 0.029 & \\
& $(n,b)$     & 0.028 & \\
& $(s,b)$     & 0.017 & \\ \hline
\end{tabular}
\end{center}
\end{table}

\begin{table}[tbp]
\caption{Equal spacing rules of Eqs. (\ref{spa}) and (\ref{spa2})
for bottom baryon masses obtained in this work (in MeV).}
\label{t11}
\begin{center}
\begin{tabular}{|c|c|}
\hline
    &  CQC  \\ 
\hline
$M_{\Sigma_b} \, + \, M_{\Omega_b}$      &   11863  \\
$2\, M_{\Xi'_b}$                         &   11878  \\ \hline
$M_{\Sigma^*_b} \, + \, M_{\Omega^*_b}$  &   11908  \\
$2\, M_{\Xi^*_b}$                        &   11922  \\ \hline
$\delta_{\Sigma_b} + \delta_{\Omega_b}$  &      45  \\
2$\delta_{\Xi_b}$                        &      44  \\ \hline
\end{tabular}
\end{center}
\end{table}

\begin{table}[tbp]
\caption{Equal spacing rule of Eq. (\ref{eqs}) for different
models in the literature. Masses are in MeV.}
\label{t10}
\begin{center}
\begin{tabular}{|c||ccccccccc|}
\hline
Mass difference  &  CQC &  
\protect\cite{Ron95} & 
\protect\cite{Sil96} &
\protect\cite{Jen96} &
\protect\cite{Bow96} &
\protect\cite{Mat02} &
\protect\cite{Liu07} &
\protect\cite{Rob07} &
\protect\cite{Ebe08} \\
\hline
$\Xi'_b(1/2^+)-\Sigma_b(1/2^+)$  & 132 & 130 & 96 & 122 & 130 &  89 &  94 & 125 & 132 \\
$\Omega_b(1/2^+)-\Xi'_b(1/2^+)$  & 117 & 110 & 93 & 111 &  90 & 104 & 133 & 123 & 128 \\
$\Xi_b(3/2^+)-\Sigma_b(3/2^+)$   & 132 & 130 & 96 & 116 & 120 &  88 &  94 & 124 & 129 \\
$\Omega_b(3/2^+)-\Xi_b(3/2^+)$   & 118 & 110 & 93 & 106 & 100 & 101 & 134 & 120 & 125  \\ \hline
\end{tabular}
\end{center}
\end{table}

\begin{table}[tbp]
\caption{Masses of charmed baryons from the present work (CQC)
and other approaches in the literature compared to experimental data (in MeV).
Those data with a question mark stand for recently measured
states whose quantum numbers are not determined and they are
confronted against the possible corresponding theoretical
state. In all cases we quote the central values, in 
Ref.~\protect\cite{Mat02} error bars are of the order of 100 MeV,
40 MeV in Ref.~\protect\cite{Ron95}.}
\label{t12}
\begin{center}
\begin{tabular}{|cc||ccccccccccc|}
\hline
State & $J^P$ & CQC & Exp.  & 
\protect\cite{Cap86} &
\protect\cite{Ron95} &
\protect\cite{Sil96} &
\protect\cite{Jen96} &
\protect\cite{Mat02} &
\protect\cite{Wan02} &
\protect\cite{Liu07} &
\protect\cite{Rob07} &
\protect\cite{Ebe08} \\
\hline
$\Lambda_c$  &  $1/2^+$ & 2285 & 2286 & 2265 & 2285 & 2285 & 2284 & 2290 &      & 2271 & 2268 & 2297  \\
             &  $1/2^+$ & 2785 & 2765?& 2775 &      & 2865 &      &      &      &      & 2791 & 2772 \\
             &  $1/2^-$ & 2627 & 2595 & 2630 &      & 2635 &      &      &      &      & 2625 & 2598  \\
             &  $1/2^-$ & 2880 & 2880?& 2780 &      & 2885 &      &      &      &      & 2816 & 3017  \\
             &  $3/2^+$ & 3061 &      & 2910 &      & 2930 &      &      &      &      & 2887 & 2874  \\
             &  $3/2^+$ & 3308 &      & 3035 &      & 3160 &      &      &      &      & 3073 & 3262  \\ 
             &  $5/2^+$ & 2888 & 2880?& 2910 &      & 2930 &      &      &      &      & 2887 & 2883  \\ \hline
$\Sigma_c$   &  $1/2^+$ & 2435 & 2454 & 2440 & 2453 & 2455 & 2452 & 2452 & 2470 & 2411 & 2455 & 2439  \\
             &  $1/2^+$ & 2904 &      & 2890 &      & 3025 &      &      &      &      & 2958 & 2864  \\
             &  $1/2^-$ & 2772 & 2765?& 2765 &      & 2805 &      &      &      &      & 2748 & 2795  \\
             &  $1/2^-$ & 2893 &      & 2840 &      & 2885 &      &      &      &      &      & 3176  \\
             &  $3/2^+$ & 2502 & 2518 & 2495 & 2520 & 2535 & 2532 & 2538 & 2590 & 2534 & 2519 & 2518  \\
             &  $3/2^+$ & 2944 & 2940?& 2985 &      & 3065 &      &      &      &      & 2995 & 2912  \\       
             &  $3/2^-$ & 2772 & 2800?& 2770 &      & 2805 &      &      &      &      & 2763 & 2761  \\ \hline
$\Xi_c$      &  $1/2^+$ & 2471 & 2471 &      & 2468 & 2467 & 2468 & 2473 &      & 2432 & 2492 & 2481  \\
             &  $1/2^+$ & 3137 & 3123?&      &      & 2992 &      &      &      &      &      & 2923  \\
             & $1/2'^+$ & 2574 & 2578 &      & 2580 & 2567 & 2583 & 2599 &      & 2508 & 2592 & 2578  \\
             & $1/2'^+$ & 3212 &      &      &      & 3087 &      &      &      &      &      & 2984  \\
             &  $1/2^-$ & 2799 & 2792 &      &      & 2792 &      &      &      &      & 2763 & 2801  \\
             &  $1/2^-$ & 2902 &      &      &      & 2897 &      &      &      &      & 2859 & 2928  \\
             &  $1/2^-$ & 3004 & 2980?&      &      & 2993 &      &      &      &      &      & 3186  \\
             &  $3/2^+$ & 2642 & 2646 &      & 2650 & 2647 & 2644 & 2680 &      & 2634 & 2650 & 2654  \\
             &  $3/2^+$ & 3071 & 3076?&      &      & 3057 &      &      &      &      & 2984 & 3030  \\ 
             &  $5/2^+$ & 3049 & 3055?&      &      & 3057 &      &      &      &      &      & 3042 \\ 
             & $5/2'^+$ & 3132 & 3123?&      &      & 3167 &      &      &      &      &      & 3123 \\ \hline
$\Omega_c$   &  $1/2^+$ & 2699 & 2698 &      & 2710 & 2675 & 2704 & 2678 &      & 2657 & 2718 & 2698  \\
             &  $1/2^+$ & 3159 &      &      &      & 3195 &      &      &      &      & 3152 & 3065 \\
             &  $1/2^-$ & 3035 &      &      &      & 3005 &      &      &      &      & 2977 & 3020  \\
             &  $1/2^-$ & 3125 &      &      &      & 3075 &      &      &      &      &      & 3371  \\
             &  $3/2^+$ & 2767 & 2768 &      & 2770 & 2750 & 2747 & 2752 & 2760 & 2790 & 2776 & 2768 \\
             &  $3/2^+$ & 3202 &      &      &      & 3235 &      &      &      &      & 3190 & 3119 \\ \hline
\end{tabular}
\end{center}
\end{table}

\begin{table}[tbp]
\caption{Possible model states and spin-parity assignments
for recently discovered charmed baryons. The 'star' indicates
radial excitations.}
\label{t30}
\begin{center}
\begin{tabular}{|c|c|}
\hline
Experimental resonance (MeV) & Model states \\ \hline
\multicolumn{2}{|c|}{ $\Lambda_c$ or $\Sigma_c$} \\ \hline
2765                         & $\Sigma_c(1/2^-)$ or ${\Lambda_c(1/2^+)}^*$  \\
2880                         & ${\Lambda_c(1/2^-)}^*$ or $\Lambda_c(5/2^+)$ \\
2940                         & ${\Sigma_c(3/2^+)}^*$                      \\
2800                         & $\Sigma_c(3/2^-)$                          \\ \hline
\multicolumn{2}{|c|}{ $\Xi_c$ or $\Xi'_c$} \\ \hline
3055                         & $\Xi_c(5/2^+)$ \\ 
3123			     & ${\Xi_c(1/2^+)}^*$ or $\Xi'_c(5/2^+)$ \\
2980                         & ${\Xi_c(1/2^-)}^*$ \\
3076                         & ${\Xi_c(3/2^+)}^*$ \\ \hline
\end{tabular}
\end{center}
\end{table}

\begin{table}[tbp]
\caption{Ground state $J^P=1/2^+$ of doubly charmed and doubly bottom
baryons from the present work (CQC) and other approaches in the literature.
Masses are in MeV.}
\label{t19}
\begin{center}
\begin{tabular}{|c|cccc|}
\hline
  & $\Xi_{cc}$ & $\Omega_{cc}$ & $\Xi_{bb}$ & $\Omega_{bb}$ \\ \hline
CQC                  &  3579    &   3697     &   10189   &  10293 \\
\protect\cite{Ron95} &  3660    &   3740     &   10340   &  10370 \\ 
\protect\cite{Sil96} &  3607    &   3710     &   10194   &  10267 \\ 
\protect\cite{Mat02} &  3588    &   3698     &   $-$     &  $-$   \\ 
\protect\cite{Rob07} &  3676    &   3815     &   10340   &  10454 \\ \hline
\end{tabular}
\end{center}
\end{table}

\begin{table}[tbp]
\caption{Excitation spectra of doubly charmed and 
doubly bottom baryons ($M(J^P)-M(1/2^+)$) from the present work (CQC) and other
approaches in the literature. Masses are in MeV.}
\label{t20}
\begin{center}
\begin{tabular}{|cc|ccccc|}
\hline
State  & $J^P$ & CQC & 
\protect\cite{Ron95} & 
\protect\cite{Sil96} & 
\protect\cite{Mat02} & 
\protect\cite{Rob07} \\  
\hline
                & $3/2^+$     &  29 &  30 &  41 &  20 &  27  \\  
                & ${3/2^+}^*$ & 312 &     & 386 &     & 238  \\  
$\Xi_{bb}$      & ${1/2^+}^*$ & 293 &     & 355 &     & 236  \\  
                & $1/2^-$     & 217 &     & 262 &     & 153  \\  
                & ${1/2^-}^*$ & 423 &     & 462 &     & 370  \\  \hline
                & $3/2^+$     &  28 &  30 &  38 &  19 &  32  \\ 
                & ${3/2^+}^*$ & 329 &     & 383 &     & 267  \\   
$\Omega_{bb}$   & ${1/2^+}^*$ & 311 &     & 359 &     & 239  \\  
                & $1/2^-$     & 226 &     & 265 &     & 162  \\  
                & ${1/2^-}^*$ & 390 &     & 410 &     & 309  \\  \hline
                & $3/2^+$     &  77 &  80 &  93 &  70 &  77  \\ 
                & ${3/2^+}^*$ & 446 &     & 486 &     & 366  \\   
$\Xi_{cc}$      & ${1/2^+}^*$ & 397 &     & 435 &     & 353  \\  
                & $1/2^-$     & 301 &     & 314 &     & 234  \\
                & ${1/2^-}^*$ & 439 &     & 472 &     & 398  \\  \hline
                & $3/2^+$     &  72 &  40 &  83 &  63 &  61  \\
                & ${3/2^+}^*$ & 463 &     & 498 &     & 373  \\    
$\Omega_{cc}$   & ${1/2^+}^*$ & 415 &     & 445 &     & 365  \\  
                & $1/2^-$     & 312 &     & 317 &     & 231  \\ 
                & ${1/2^-}^*$ & 404 &     & 410 &     & 320  \\  \hline
\end{tabular}
\end{center}
\end{table}

\end{document}